\documentclass[preprint,12pt]{elsarticle}

\usepackage{amssymb}
\usepackage{amsmath}

\journal{Nuclear Physics A}

  \newcommand {\nc} {\newcommand}
  \nc {\beq} {\begin{eqnarray}}
  \nc {\eeq} {\nonumber \end{eqnarray}}
  \nc {\eeqn}[1] {\label {#1} \end{eqnarray}}
  \nc {\eol} {\nonumber \\}
  \nc {\eoln}[1] {\label {#1} \\}
  \nc {\ve} [1] {\mbox{\boldmath $#1$}}
  \nc {\ves} [1] {\mbox{\boldmath ${\scriptstyle #1}$}}
  \nc {\mrm} [1] {\mathrm{#1}}
  \nc {\half} {\mbox{$\frac{1}{2}$}}
  \nc {\thal} {\mbox{$\frac{3}{2}$}}
  \nc {\fial} {\mbox{$\frac{5}{2}$}}
  \nc {\la} {\mbox{$\langle$}}
  \nc {\ra} {\mbox{$\rangle$}}
  \nc {\etal} {\emph{et al.}}
  \nc {\eq} [1] {(\ref{#1})}
  \nc {\Eq} [1] {Eq.~(\ref{#1})}
  \nc {\Sec} [1] {Sec.~\ref{#1}}
  \nc {\chap} [1] {Chapter~\ref{#1}}
  \nc {\anx} [1] {Appendix~\ref{#1}}
  \nc {\tbl} [1] {Table~\ref{#1}}
  \nc {\Fig} [1] {Fig.~\ref{#1}}
  \nc {\ex} [1] {$^{#1}$}
  \nc {\Sch} {Schr\"odinger }
  \nc {\flim} [2] {\mathop{\longrightarrow}\limits_{{#1}\rightarrow{#2}}}
  \nc {\IR} [1]{\textcolor{red}{#1}}
  \nc {\IB} [1]{\textcolor{blue}{#1}}
  \nc{\IG}[1]{\textcolor{green}{#1}}

\begin{document}

\begin{frontmatter}



\title{First Experimental Test of the Ratio Method}


\author[bnl]{S.\ Ota} 
\author[jgu]{P.\ Capel\corref{cor1}} 
\ead{pcapel@uni-mainz.de}
\author[stm]{G.\ Christian} 
\author[jgu]{V.\ Durant}
\author[tamu]{K.\ Hagel} 
\author[tamu]{E.\ Harris}
\author[surrey]{R.C.\ Johnson}
\author[tamu]{Z.\ Luo}
\author[msu]{F.M.\ Nunes}
\author[tamu]{M.\ Roosa}
\author[tamu]{A.\ Saastamoinen}
\author[tamu]{D.P.\ Scriven}

\affiliation[bnl]{organization={National Nuclear Data Center, Brookhaven National Laboratory},
            city={Upton},
            postcode={11973-5000}, 
            state={NY},
            country={USA}}
            
\affiliation[jgu]{organization={Institut fuer Kernphysik, Johannes Gutenberg-Universitaet Mainz},
            city={Mainz},
            postcode={55099}, 
            country={Germany}}

\affiliation[stm]{organization={Department of Astronomy \& Physics, Saint Mary's University},
            city={Halifax},
            postcode={B3H~3C3}, 
            state={NS},
            country={Canada}}

\affiliation[tamu]{organization={Cyclotron Institute, Texas A\&M University},
            addressline={}, 
            city={College Station},
            postcode={77843}, 
            state={Texas},
            country={USA}}

\affiliation[surrey]{organization={Department of Physics, University of Surrey},
            city={Guildford },
            postcode={GU2 7XH}, 
            country={United Kingdom}}

\affiliation[msu]{organization={Facility for Rare Isotope Beams, Michigan State University},
            city={East Lansing},
            postcode={48824}, 
            state={MI},
            country={USA}}


\begin{abstract}
The ratio is a new reaction observable suggested to extract accurately structure information on halo nuclei.
It corresponds to the ratio of differential cross sections for scattering and breakup, which is predicted to remove the uncertainty related to the reaction dynamics.
We present here the first experimental test of the method for the $^{11}$Be + $^{12}$C collision at $E_{\rm Lab}=20A$ MeV performed at Texas A\&M University.
Differential cross sections for scattering and inclusive one-neutron breakup have been measured with the new detector array BlueSTEAl.
The ratio of cross sections is very smooth and independent of the projectile-target interaction, which demonstrates the validity of the \emph{ratio method}.
We extend our analysis to existing $^{11}$Be + $^{208}$Pb data, confirming that the method works well on any target.
\end{abstract}




\end{frontmatter}



The production of radioactive-ion beams in the mid-80s has enabled the study of nuclei far from stability.
This technical breakthrough has led to the discovery of exotic nuclear structures such has \emph{halo nuclei}.
These light, neutron-rich nuclei exhibit a large matter radius.
This unusual size is due to the small binding energy of one or two neutrons.
Thanks to this loose binding the valence neutrons can tunnel far away from the other nucleons and form a diffuse halo around a compact core \cite{Tan96}.
Examples are $^{11}$Be with one neutron in its halo and the two-neutron halo nucleus $^{11}$Li.
This strongly clusterised structure challenges modern nuclear models, and is therefore the focus of many theoretical and experimental studies.

Because halo nuclei are very short lived, most of the information gathered about their structure is obtained through reactions.
In breakup, the halo is dissociated from the core during the collision with a target \cite{NK12}.
This clearly reveals halo structures.
Unfortunately, the detailed analysis of experimental data is hampered by the sensitivity of breakup calculations to the optical potentials, which account for the interaction between the projectile constituents and the target \cite{Capel2004}.
The structure information inferred from experiment seems inherently marred by the uncertainty of the input of the reaction models.

A detailed analysis of differential cross sections for elastic scattering and breakup of halo nuclei has shown that both processes exhibit very similar diffraction patterns \cite{CHB10}, indicating that the projectile is scattered in a similar way whether it remains bound or if it dissociates.
This is easily explained within the Recoil Excitation and Breakup model (REB) \cite{JAT97}.
In that model, the scattering cross section factorises as
\beq
\frac{d\sigma_{\rm el}}{d\Omega}&=&|F_{00}|^2\ \left(\frac{d\sigma}{d\Omega}\right)_{\rm pt},
\eeqn{e1}
where $\left(\frac{d\sigma}{d\Omega}\right)_{\rm pt}$ is the scattering cross section computed for a pointlike projectile, 
while the form factor $F_{00}=\int |\Phi_0(\ve{r})|^2e^{i\ve{Q\cdot r}}d\ve{r}$ accounts for the extension of the halo wave function $\Phi_0$; $\ve{Q}=\frac{m_{\rm n}}{m_c+m_{\rm n}}(\ve{K}-\ve{K'})$ is proportional to the exchanged momentum.
A similar factorisation is obtained for the breakup cross section, but for the form factor, which includes also continuum wave functions.
However, the cross section $\left(\frac{d\sigma}{d\Omega}\right)_{\rm pt}$ appearing in the REB factorisations is the same for scattering and breakup.
This explains the similarity of the diffraction patterns noted in Ref.~\cite{CHB10}.

In addition, this explanation brings a new idea.
Taking the ratio of these cross sections should remove the strong dependence on the reaction process and on the projectile-target interaction.
The REB predicts also that this ratio will be function of form factors of the projectile wave functions, which depend only on the projectile structure.
Moreover, corresponding to the ratio of cross sections, it does not require a precise normalisation of the cross sections, a precious experimental advantage.
This new reaction observable therefore provides a very accurate tool to study the structure of halo nuclei.

This idea has been studied theoretically by comparing accurate reaction calculations to the REB predictions \cite{Capel2011,Capel2013}.
These tests show the interest and accuracy of the ratio method to study halo nuclei.
By removing the dependence on the reaction mechanism, the ratio enables a direct and more accurate access to the projectile structure.

\begin{figure}[t]
\centering
\includegraphics[width=0.466\linewidth]{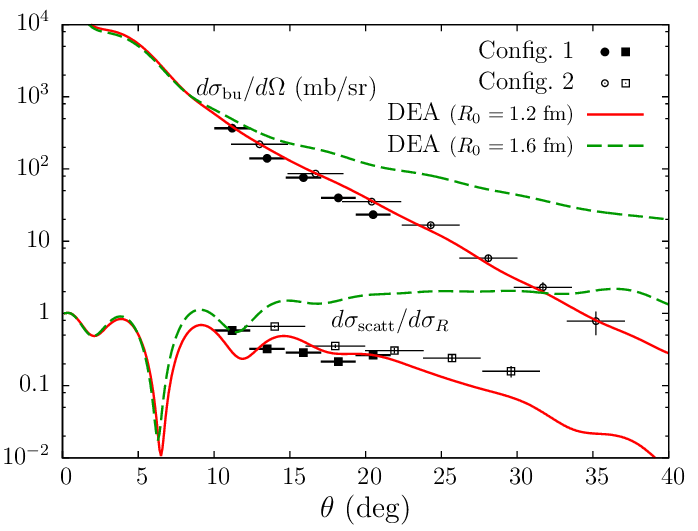}\hfill 
\includegraphics[width=0.50\linewidth]{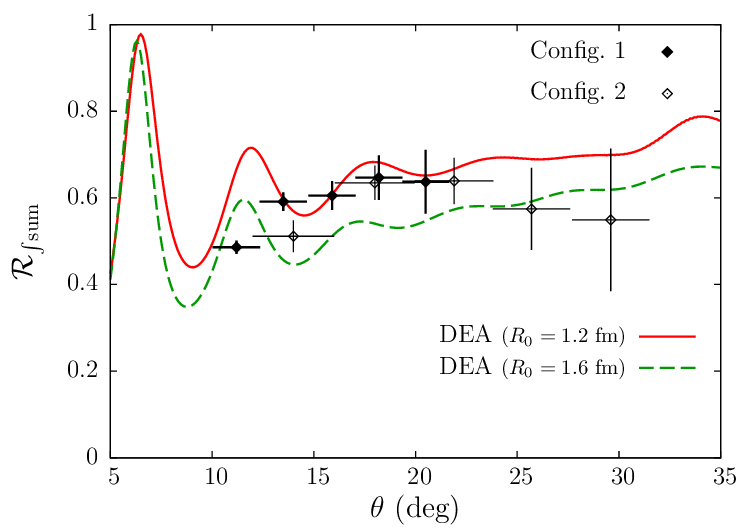}
\caption{First experimental study of the ratio method in the collision of $^{11}$Be on C at $22.8A$~MeV \cite{OC24}.
Left: differential cross sections for scattering (squares, ratio to Rutherford) and inclusive breakup (circles, in mb/sr) compared to reaction calculations performed with two different choices of optical potentials (red solid and green dashed lines).
Right: The ratio exhibits a smooth behaviour and is reproduced with both optical-potential choices.
}\label{f1}
\end{figure}

In this work, we test the ratio method experimentally for the first time \cite{OC24}.
The measurement took place using the K500 superconducting cyclotron of Texas A\&M University.
A $^{11}$Be beam was produced by bombarding a $^9$Be target with a $30A$~MeV $^{13}$C primary beam.
After fragment separation, a $^{11}$Be beam of about 60\% purity at $22.8A$~MeV was directed on a C$_{\rm nat}$ target.
The $^{11}$Be scattering and $^{10}$Be breakup charged fragment after collision were detected with the new BlueSTEAl detector system \cite{Ota2023}.
The corresponding differential cross sections are shown in \Fig{f1} (left).
They both exhibit a significant angular dependence, varying by about three orders of magnitude over the measured angular range.
As predicted theoretically, this angular dependence is removed in their ratio displayed in \Fig{f1} (right), leading to a smooth observable, which confirms its independence of the reaction  process.

The individual experimental cross sections shown in \Fig{f1} (left) are well reproduced by a dynamical calculation with an appropriate choice of $^{10}$Be-C optical potential (red solid lines).
Changing that interaction leads to a clear disagreement with the measurement (green dashed lines).
However, both choices reproduce the experimental ratio, see \Fig{f1} (right).
Contrary to individual cross sections, the ratio is independent of optical potentials, a clear advantage in the study of exotic nuclei far from stability, where optical potentials are not well constrained.
Similar results have been obtained in a reanalysis of the collision of $^{11}$Be on Pb at $19.1A$~MeV measured by Duan \etal\  \cite{Duan2022}.
This confirms that the method is independent of the target used.

The ratio method provides an accurate and model-independent way to study exotic nuclei far from stability \cite{Capel2011,Capel2013}.
This first experimental study confirms the theoretical predictions \cite{OC24}.
We plan to exploit the high beam intensity of FRIB to use the ratio method to study the structure of $^{19}$C and other, heavier, halo nuclei.




  \bibliographystyle{elsarticle-num} 
  \bibliography{11Be_Manuscript}






\end{document}